# Ultrafast atomic dimerization of Peierls distortion in semimetal molybdenum ditelluride

Zhong Wang(王众)[1,2], Chunlong Hu(胡春龙)[1,2], Changchang Gong(巩畅畅)[1,2], Fuyong Hua(华傅甬)[1,2], Qian You(游骞)[1,2,†], Wenxi Liang(梁文锡)[1,2,*]

[1]Wuhan National Laboratory for Optoelectronics, Huazhong University of Science and Technology, 1037 Luoyu Road, Wuhan 430074, China

[2]Advanced Biomedical Imaging Facility, Huazhong University of Science and Technology, Wuhan 430074, China

†Current address: Zhongshan Lighting Fast Intellectual Property Rights Service Center, Zhongshan 528421, China

*Corresponding author. Email: wxliang@hust.edu.cn

Semimetal molybdenum ditelluride (1T′-$MoTe_2$) possess diverse phase transitions enriching its application prospects. The structural response during these transitions is crucial to understanding the underlying mechanisms, but the desired details of pathway and time span are still insufficient. Here, we investigate the lattice evolution in few-layer 1T′-$MoTe_2$ after photoexcitation, using ultrafast electron diffraction and density functional theory (DFT) calculations. The observed complex lattice responses with unintuitively evolving Bragg peak intensity and interplanar spacing, are best interpreted as the combination of shear displacement and Mo-Mo bond shortening in a few picoseconds, and a metastable structure in nanoseconds, basing on the analyses of structure factor and pair distribution function. The DFT calculations reveal that, the photodoped electrons induced population change of the antibonding states close to Fermi level, lead to the shear displacement and the dimerization of Mo pairs. Our findings present new insights for elucidating the picture of Peierls distortion in 1T′-$MoTe_2$.

*Introduction.* The trigonal prismatic (2H) and octahedral (1T) coordinates are two typical structural phases of transition metal chalcogenides (TMDCs),[1] thus the transitions between different phases render TMDCs abundant potentials in applications and the studies of physical property. Molybdenum ditelluride ($MoTe_2$) with layered lattice structure is an attractive member of the big TMDCs family, serving as a prototype example. $MoTe_2$ turns into the monoclinic 1T′ phase from the unstable 1T phase through Peierls distortion, which is characterized by the Mo-Mo metallic bonds.[2] As a semimetal, 1T′-$MoTe_2$ shows potentials in nonlinear optics due to its in-plane anisotropy.[3] Moreover, the theoretical calculation[4, 5] predicted that 1T′-$MoTe_2$ is a candidate for 3D second-order topological insulators due to the band inversion, suggesting possible applications in spintronic devices and quantum computing.[6] The inversion symmetry of 1T′-$MoTe_2$ breaks at low-temperature (below ∼250 K for bulk specimen), then the orthorhombic $T_d$ phase forms through tilting the axis along the out-of-plane direction.[7] As a Weyl semimetal, $T_d$-$MoTe_2$ provides an ideal platform for the studies of negative magnetoresistance, anomalous quantum Hall effect, and chiral magnetic effects.[4]

In the theoretical descriptions, the hole and electron doping stabilize the 1T′ and $T_d$ phases, respectively.[7] Such transitions between two phases by doping at the Mo sites were experimentally demonstrated.[8, 9] However, the monoclinic-to-orthorhombic transition is expected to vanish in the morphology of thin flake (typically less than 20 nm), because the exfoliation method or the exposure to atmospheric oxygen and moisture brings spontaneous hole doping.[6] In terms of experiment, the optical probes provide the access of manipulating structural phases in $MoTe_2$, as the laser pulses dope electrons into the conduction band or generate a large enough electric field, therefore changing the landscape of free energy and inducing a phase transition. The ultrafast phase transitions in $MoTe_2$ using optical pump-probe reflectivity measurements have been reported,[10, 11] in which the 0.42 THz shear mode was regarded as an indicator of the existence of $T_d$ phase, because this phonon mode is Raman inactive in the centrosymmetric 1T′ phase.

On the microscopic level, the topological band inversion or the appearance of Weyl nodes is induced by the intralayer Peierls distortion or the interlayer shear distortion,[12] therefore the probe with lattice sensitivity is indispensable in elucidating these processes. For example, the picosecond-level symmetry switch in $WTe_2$ was monitored by mega-electron-volt ultrafast electron diffraction (UED);[13] The concurrent shear displacement and Peierls distortion suppression in $MoTe_2$ were recently identified by UED as well.[12] In these reports, the calculation of structure factor, which is primarily determined by the fractional coordinates of the atoms in unit cell, is the main approach to model the lattice distortion accounting for the diffraction intensity change as a function of atomic displacement, although some structural details are ignorable in certain cases.[14] The picture of light-driven lattice motions during topological phase transition is still elusive, further understanding of these processes is desired for understanding the quantum phenomena in TMDCs, e.g., Weyl fermions.[15]

Here, we investigate the photoexcited structure relaxation in few-layers 1T′-$MoTe_2$, employing the UED measurement and density functional theory (DFT) calculation. The intensity and position changes of diffraction peak measured by UED reveal that, besides the Debye-Waller effect, the shear displacement and the dimerization of Mo atoms occur within several picoseconds. These ultrafast structure disorders are explained by the distribution characteristics of electronic band, electronic density of states (DOS), and crystal orbital Hamilton population (COHP) obtained through the DFT calculations around Fermi level. Furthermore, the intensity drops on nanosecond timescale, which deviates from the Debye-Waller model, suggest a long-lasting transient structure under high excitation power. Our findings construct the scenario of lattice responses correlated to the Peierls distortion in photodoped 1T′-$MoTe_2$.

*Results and discussions.* The investigated 1T′-$MoTe_2$ film with thickness of ~7 nm (~10 layers) was synthesized by chemical vapor deposition (see Methods in the Supplementary Information), then transferred onto a bare copper grid as the free-standing film. The sample was grown with the morphology of quasi-single-crystal, in which the grains possess lateral sizes of tens to hundreds of nanometers, while the *c*-axis of all grains aligns along the direction perpendicular to the film surface, see Fig.1(a). The Raman spectrum with 3 characteristic peaks[16] and the selected area electron diffraction with sharp spots and are shown in Fig.1(b), confirming the 1T′ structure. The schematic layout of UED is illustrated in Fig.1(c), showing the diffraction patterns which encode the transient lattice structure under excitation of 800 nm (see Methods in the Supplementary Information). A radially averaged distribution of diffraction intensity through Hough transform is shown in Fig.1(d).

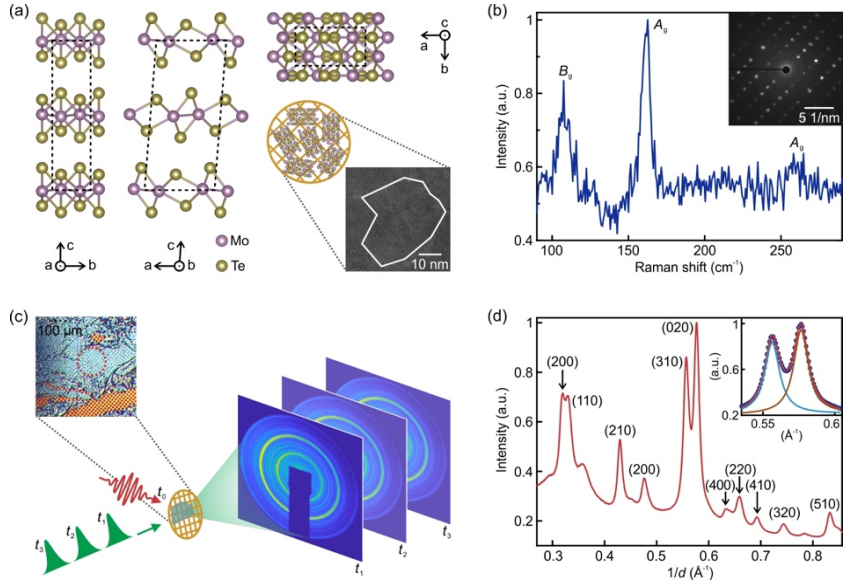

**Fig. 1.** Sample characterization, lattice structure, and UED measurement. (a) Monoclinic lattice of 1T'-MoTe$_2$ with layered structure bonding by van der Waals force. Dashed frames, unit cell. Lower right panel, high-resolution transmission electron microscopy image with a crystalline grain outlined in white, and the schematic of grain orientations. (b) Raman spectrum. Inset, selected area electron diffraction pattern. (c) Schematic layout of UED measurement, with estimated temporal resolution less than 1 ps.[17] Inset, optical photograph of the sample, marked with the probed area (dotted circle). (d) Radially averaged distribution of diffraction intensity. Inset, Lorentzian fit of the (310) and (020) peak profile.

Upon the femtosecond pulse excitation, the 1T′-MoTe$_2$ shows complex structural response, as depicted in Fig. 2(a), The response exhibits neither straightforward intensity drop of Bragg peaks, nor intensity drops scaling with the scattering vector as the description of Debye-Waller model.[18] As depicted in Fig. 2(b), the intensity drops extracted at the delay times of 3 ps and 3 ns show that the decay processes deviate from the linear trend of Debye-Waller effect,[19] which is contributed by the increase of atomic mean square displacement (MSD). Such deviations suggest the occurrence of nonthermally driven transient lattice distortions on the picosecond and nanosecond scales.

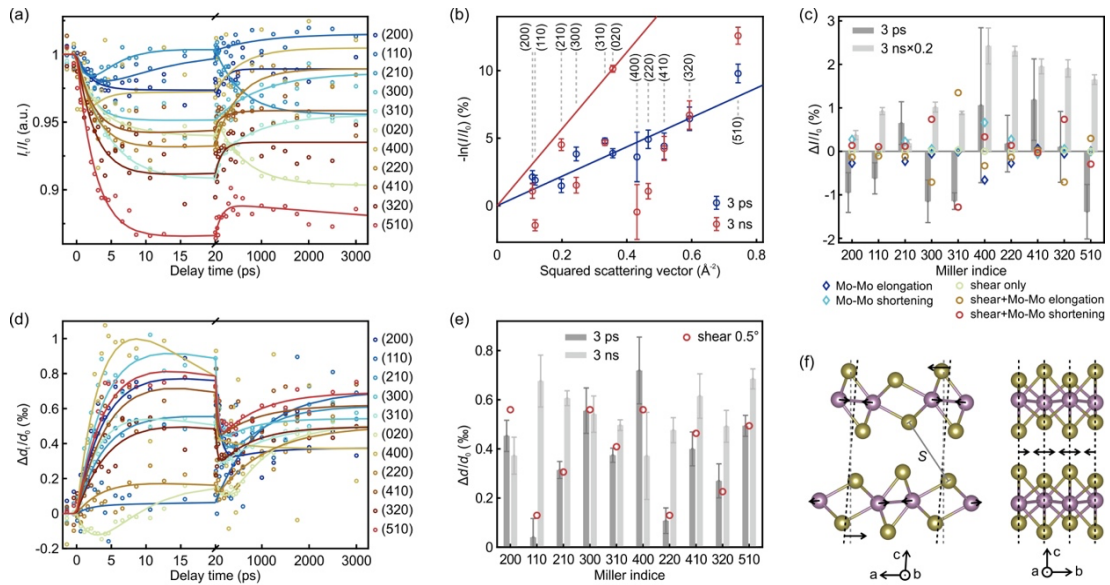

**Fig. 2.** Lattice responses measured by UED and the nonthermal lattice distortions. (a) Intensity drops of Bragg peaks upon excitation of 1.2 mJ/cm$^2$. Solid lines, multiexponential fits. (b) Nonlinear dependence of the intensity drops at 3 ps and 3 ns on the squared scattering vector. Both solid lines across the zero point and the value of (020) peak. (c) Nonthermal intensity changes (grey bars, with the contribution of thermal motion subtracted) and the intensity changes by the five distortion models (color symbols) calculated with atomic displacement of ~1‰. Error bars, the standard deviation of measurements. (d) Kinetics of interplanar spacing change. (e) Measured interplanar spacing changes (grey bars) and the calculated changes with shear displacement (red circles). (f) Schematic of the lattice distortion from tilted structure towards orthogonal structure induced by the shear displacement (left), and from the Mo-Mo bond shortening (right).

The lattice response recorded in electron diffraction is given by the structure factor[20] as:

$$F(hkl) = \sum_j f_j \exp\left[-i2\pi(hx_j + ky_j + lz_j)\right], \quad (1)$$

where the summation runs over all atoms in the unit cell, $f_j$ and $(x_j, y_j, z_j)$ are the atomic scattering factor and the fractional coordinates of the $j$th atom, respectively, and $(hkl)$ are the Miller indices. So far the interlayer shear displacement along the $a$-axis[10] and the intralayer Mo-Mo metallic bonds elongation[12] are two well adopted candidates for the origin of the fractional coordinate changes upon photoexcitation. We calculated the diffraction intensity changes induced by the distortion models combining these two motions through equation (1), then compared the results to the measured intensity changes.

Given that the structure factor of (020) crystal plane is not affected by these two motions, the intensity drop of (020) peak is considered constituting only of the contribution of increased MSD, which is better demonstrated by the measurement with varied fluences, as discussed later. Therefore, we take the (020) intensity drop as a calibration to estimate the contribution of nonthermal distortions to the intensity drop of other peaks, i.e., the intensity drop of all other Bragg peaks is assumed to constitute of the contributions combing the thermal MSD effect scaled with that of (020) peak, and the nonthermal distortions. Taking the intensity change of (310) peak as an example, which shows the best signal-to-noise, only the combination of shear displacement and Mo-Mo bond shortening gives the proper intensity drop at 3 ps when the component of MSD contribution is subtracted. However, no candidate among all the considered distortion model yields a result satisfying the intensity changes at 3 ps for all measured Bragg peaks, as depicted in Fig. 2(c). A potential cause is the large number of defects in the very thin sample, in which the defect induced local distortions, and polarons as well,[21, 22] possibly make significant contributions. At the long delay time of 3 ns, all other peaks show intensity increases when the MSD contributions are subtracted, as depicted in the upper part of Fig. 2(c). Such diffraction intensity evolutions are possibly interpreted as the increase of atomic displacement in $b$-$c$ plane, e.g., the formation of diamond-shaped metallic bonds in other TMDCs,[23, 24] which brings additional intensity drop to the (020) peak, thus resulting in the relative intensity increase in other peaks.

Although the intensity changes show elusive structural evolutions, the lattice responses encoded in the change of interplanar spacing give a clearer picture. Fig. 2(d) depicts the kinetics of relative interplanar spacing extracted from the measured shift of Bragg peaks, showing different expansions for all planes except (020) at the early 10 ps, followed by the quite uniform expansions for all planes on the nanosecond scale. At the delay time of ~3 ps, the transient expansion is more significant for the lattice planes with smaller angle with respect to the $b$-$c$ plane, agreeing with the trend of distortion induced by the shear displacement, which expands the interplanar spacings except the (020) plane, as illustrated in Fig. 2(f). We calculated the changes of interplanar spacing with the shear displacement model, finding that the results agree well with the experimental observations at 3 ps when the angle (termed as $\beta$) between the $a$- and $c$-axis decreases ~0.5°, as depicted in Fig. 2(e). The reduced $\beta$ is indicative of the phase transition from 1T′ to T$_d$ (a decrease of ~4° for completing the transition), which is facilitated by doping electron via photoexcitation.[11] However, no complete transition was observed in our measurement, due probably to the holes, which stabilized the monoclinic structure,[6] in the thin sample (with thickness less than 20 nm) introduced during the exposure to air. In contrast to the expansion of other planes, the separation of (020) planes shrinks rapidly in the early few picoseconds, agreeing with the model of Mo-Mo bond shortening, which projects interplanar spacing decrease along the *b*-axis, as illustrated in Fig. 2(f). At the long delay time of nanosecond, all lattice planes tend to expand uniformly, showing the feature of thermal expansion, as expected. The uniform expansion in turn manifests that the nonthermal distortions only occur within the lattice point, i.e., changing the fractional coordinates.

Then the lattice distortions are further revealed through the analysis of radial pair distribution function (PDF),[25] which is an effective approach to extract the details of atomic motion proven in many UED studies.[26, 27] In our case of quasi-single-crystal sample, only the changes of atom pair separation projected in the *a-b* plane is accessible, as the UED probes along the *c*-axis. We therefore constructed a structure with reduced diffraction elements to simplify the PDF calculation. The elements are constructed with the positions of atoms in the same column projected in the *a-b* plane, as illustrated in Supplementary Fig. S1. Thus, their scattering factors can be written as:

$$f_{\text{Mo,Te}} = L \times f_{\text{Mo,Te}}, \qquad (2)$$

where $L$ is the number of layer in the sample, and $f_{\text{Mo,Te}}$ denotes the scattering factors of Mo and Te atoms, which are estimated by the empirical parameter.[28] The calculated PDF (see Supplementary Note 1) before excitation is depicted in Fig. 3(a), showing broad peaks with spatial resolution of ~1 Å. The limited PDF resolution is originated from the combination of the instrument camera length, the specifications of detector, and the coherence of probe electrons, resulting in no single atom pair separation resolved.

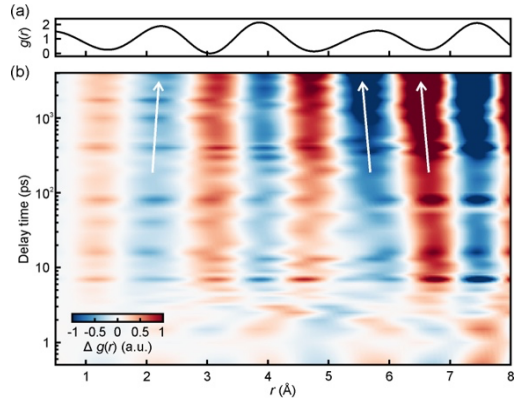

**Fig. 3.** PDF analysis of atomic displacement. (a) Static PDF. (b) Pseudocolor contour plot of the evolution of differential PDF upon excitation of 1.2 mJ/cm$^2$. Arrows, the guides to the eye.

Considering the first peak is contributed by multiple reduced diffraction elements almost overlapping each other in the *a-b* plane, here we only discuss qualitatively the evolutions of atomic displacement encoded in the temporal resolved PDF. After excitation, the PDF evolves with three stages, as shown in Fig. 3(b). First, the peaks rise and the valleys decrease slightly at the early ~3 ps, contradicting the behaviors of lattice thermalization. Instead, such PDF evolution suggests the nonthermal lattice distortions, which can be induced by local distortions, for example, the bond shortening induced by the dimerization of Mo atoms. In the next stage, which lasts several tens of picoseconds, the PDF evolves in an opposite trend with peak decrease and valley increase, suggesting the dominant roles of thermally driven atomic vibration and lattice expansion. In the last stage, which starts from ~200 ps, the PDF evolves in the same trend as the previous stage, but with peak and valley shifts as indicated by the arrows in Fig.3(b), suggesting a possible metastable structure on nanosecond scale.

The three-stage evolution of PDF is testified by the UED measurements using various excitation fluences, which also demonstrate the three-stage evolutions of intensity drop and interplanar spacing change, as depicted in Fig. 4, taking (020) peak as an example. Upon excitations with fluence more than 1 mJ/cm$^2$, the (020) intensity drop is nearly saturated during the first two stages, but apparently increased after ~200 ps. The same fashion applies to the interplanar spacing change. As the excitation fluence increase, i.e., when the photodoped carriers increase, both the intensity drop and the interplanar spacing change at 3 ns deviate from the linearity of lattice thermalization, as depicted in the inset of Fig. 4, again suggesting a long-lasting metastable structure. Although these observations give no direct evidence of lattice distortions, the evolution trends corroborate the inferred picture of structural responses: The transient distortions, including the shear displacement and the dimerization of Mo atoms, take place in a few picoseconds accompanied with lattice thermalization, then relax in several tens of picoseconds; Upon excitation with high fluence, the system possibly reaches a metastable phase with other distortion after ~200 ps,[26] alongside the uniform lattice expansion, lasting to the nanosecond time scale.

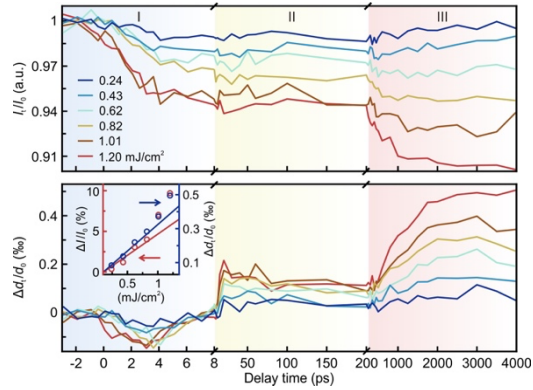

**Fig. 4.** Intensity drop and interplanar spacing change of (020) peak upon excitation of various fluences. Color shades highlight the three stages of evolution. Inset, the dependence of intensity drop and interplanar spacing change at 3 ns on excitation fluence; Solid lines, the guides to the eye.

Given that the shear mode driven by electron-phonon coupling induced the change of potential energy surface and the subsequent interlayer structural transition in 1T′-MoTe$_2$[12], we computed the electronic band structure using the Vienna Ab initio Simulation Package code[29] (see Methods in the Supplementary Information). The computation was implemented with an on-site Coulomb repulsion (U), which was set to 5.0 eV for the Mo 4$d$ orbitals,[7] to account for the prominent effects of many-body interaction[2] such as Peierls distortion. The obtained band structure is shown in Fig. 5a, in which we found that a partially occupied valence band crosses the Fermi level along the $\Gamma$-$A$ direction. This valence band is mostly composed of the $p$ orbitals of Te, and related to the interlayer antibonding states. After photoexcitation, the antibonding states are more occupied by electrons, and the separation between two Te atoms in the adjacent layers (labeled as $S$, see Fig. 2(f)) elongate to drive the occurrence of shear distortion.[7] Furthermore, the small DOS around the Fermi level, as shown in Fig. 5(b), facilitates the slowing down of hot electron relaxation,[30, 31] thus resulting in the long carrier lifetime similar to that of semiconductor phase,[32] and the enhancement of shear distortion. However, the holes in the sample probably capture the injected electrons therefore stabilize the 1T′ phase, suppressing the shear distortion and the full transition towards the T$_d$ phase.

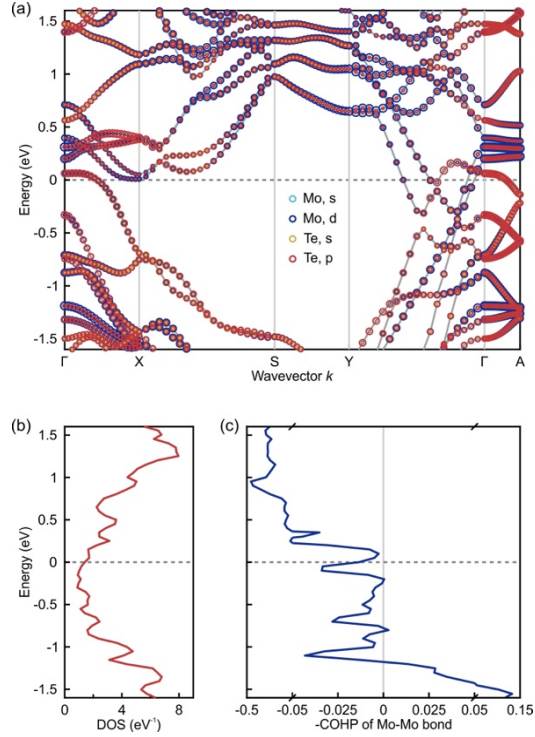

**Fig. 5.** Calculated band structure, DOS, and COHP. (a) Orbital-projected electronic band. The size of marker denotes the contribution strength of each orbital. The *s* orbitals of Mo are mainly located below -1.5 eV, thus hardly visible. (b) DOS of bulk 1T′-MoTe$_2$. (c) COHP of Mo-Mo bond. The *d* orbitals of Mo with energy higher than -1.15 eV are antibonding states. Dashed line, the Fermi Level.

We also calculated the COHP, which indicates the bonding and antibonding contributions to the band-structure energy, using the LOBSTER program.[33] The COHP between the nearest Mo atoms is depicted in Fig. 5(c), showing positive value of COHP-Mo above the Fermi level. The positive COHP-Mo indicates that all excited electrons are occupying the antibonding *d* orbitals of Mo,[34] resulting in the elongation of Mo-Mo metallic bond length, which takes place in a few hundred femtoseconds in reported work.[12] Our observation of Mo-Mo bond shortening is probably induced by two factors. First, the local maximum at ~-0.1 eV and minimum at ~0.1 eV of COHP around the Fermi level, together with the retarded relaxation of hot electrons induced by the small DOS, lead to the decrease of integral COHP of the electron occupied states during the first few picoseconds upon weak excitation (~0.015 e/cell in our study, which is much smaller than the ~0.4 e/cell for Mo-Mo elongation). Second, the temporal resolution of our UED facility is insufficient to resolve the rapid intraband relaxation (less than 500 fs in MoTe$_2$[35, 36]), recording only the states after carrier relaxation. The holes in bonding states and the electrons in antibonding states relaxing to the *d* orbitals of Mo around the Fermi level, are discernable and recorded in our study.

*Conclusion*. In summary, we observed the transient lattice distortions in the photoexcited thin film of quasi-single-crystal 1T′-MoTe$_2$, which showed the evolutions of diffraction features out of the straightforward lattice thermalization described the Debye-Waller model. Combining the structure analyses and DFT calculations, these behaviors can be interpreted as the interlayer shear displacement, and the intralayer Mo-Mo dimerization which is correlated to the enhancement of Peierls distortion. The Peierls transition and topological transition in MoTe$_2$ and other materials with alike properties, are originated from the spontaneous symmetry breakings towards a lower system energy, which is possibly destroyed by external disturbances. Such disturbing in turn feeds back the information of the system, for instance, melting the dimers[14] and annihilating the Weyl fermions[10] with photoexcitation enable the study of involved ultrafast processes. Likewise, our results bring the lattice responses of 1T′-MoTe$_2$ upon photoexcitation, as well as demonstrating the possibility of using this approach to drive the Peierls distortion, and probably further manipulating these phase transitions.

**Acknowledgements**

We thank Dr. Zewen Xiao for the helps on DFT calculations.

**Author contributions**

W.L. conceived of and supervised the project. Z.W. performed the measurements with supports from C.H., C.G., F.H. and Q.Y. Z.W. and W.L. analyzed the data with discussions with all authors. Z.W. and W.L. wrote the manuscript with contributions from all authors.

**Competing interests**

The authors declare no competing interests.

# Supplementary Information

# Ultrafast atomic dimerization of Peierls distortion in semimetal molybdenum ditelluride

Zhong Wang(王众)[1,2], Chunlong Hu(胡春龙)[1,2], Changchang Gong(巩畅畅)[1,2], Fuyong Hua(华傅甬)[1,2], Qian You(游骞)[1,2,†], Wenxi Liang(梁文锡)[1,2,*]

[1]Wuhan National Laboratory for Optoelectronics, Huazhong University of Science and Technology, 1037 Luoyu Road, Wuhan 430074, China

[2]Advanced Biomedical Imaging Facility, Huazhong University of Science and Technology, Wuhan 430074, China

†Current address: Zhongshan Lighting Fast Intellectual Property Rights Service Center, Zhongshan 528421, China

*Corresponding author. Email: wxliang@hust.edu.cn

**Methods**

**Sample preparation.** The 1T′-MoTe$_2$ film (~10 layers) was deposited on a Si wafer by the chemical vapor method (SixCarbon Technology, Shenzhen). After etching the Si substrate by hydrogen fluoride, the 1T′-MoTe$_2$ film was floated on the liquid surface. Then a bare transmission electron microscopy grid was used to attach the thin films as the final sample.

**Ultrafast electron diffraction.** The UED measurements were performed under the ultrahigh vacuum condition, using a home-built ultrafast electron diffractometer with estimated temporal resolution of subpicosecond.[1] A Ti:sapphire regenerative amplifier (Legend, Coherent) delivering 40 fs pulses of 800 nm with repetition rate of 5 kHz, was used to excite the sample and stimulate the photoelectron packs for capturing the transient structure of the excited sample. The pulsed probe electrons were accelerated by an electric field of 30 kV, generating diffraction patterns through a transmission geometric setup. The diffraction signals were gained by two chevron-stack microchannel plates, then recorded by a CMOS camera (ORCA-Flash, Hamamatsu). Each pattern was recorded with 30000 accumulated pulses. The diameter of pump and probe spots were ∼700 μm and ∼150 μm, respectively.

**Density functional theory (DFT) calculation.** The calculations of geometry optimization, electronic band structure, and density of states were performed using the Vienna *Ab initio* Simulation Package (VASP) code[2] with the generalized gradient approximation of Perdew–Burke–Ernzerhof (GGA-PBE). The projected augmented wave method was utilized with a plane-wave basis set,[3] with the convergence criteria set to 10⁻⁵ eV for energy and 0.001 eV/Å for force. A kinetic cutoff energy of 350 eV and a Monkhorst-Pack[4] special k-point mesh of 14×14×3 were employed for the calculations. An on-site Coulomb repulsion (U) was set to 5.0 eV[5] for the Mo 4$d$ orbitals. Additionally, the DFT-D3 method developed by Grimme was employed to account for the van der Waals interactions.[4] The crystal orbital Hamilton population (COHP) was obtained from the DFT using the LOBSTER program,[6] and the calculated absolute charge spilling is 1.48%.

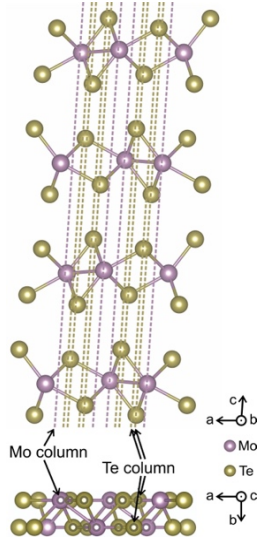

**Figure S1.** Schematic of the construction of reduced diffraction elements including 12 kinds of atom columns.

## Supplementary Note 1: Calculations of pair distribution function (PDF)

The PDF is calculated on the basis of a reduced distribution function $G(r)$, which can be obtained by the sine transform of the distribution of electron diffraction intensity:[7, 8]

$$G(r) = \frac{2}{\pi}\int_{Q_{\min}}^{Q_{\max}} Q[S(Q) - 1]\sin(Qr)\,e^{-0.03Q^2}dQ, \qquad (1)$$

where $Q$ is the scattering vector, $r$ represents the separation between the reduced diffraction elements (see Figure S1) in real space, $S(Q)$ is the structure function of the sample, and $e^{-0.03Q^2}$ is a damping term to reduce the edge effects in transform. After the background subtraction, the relationship between $S(Q)$ and the measured coherent scattering intensity $I(Q)$ is expressed as:

$$S(Q) = 1 + \frac{N \cdot I(Q) - \langle f_e^2(Q)\rangle}{\langle f_e(Q)\rangle^2}, \qquad (2)$$

where $f_e(Q)$ is the electron scattering factor of the reduced diffraction elements.[9] The parameter $N$ is a normalization factor depending on the scattering factor and intensity, which is given by

$$N = \frac{\int_{Q_{\min}}^{Q_{\max}}\langle f_e^2(Q')\rangle dQ'}{\int_{Q_{\min}}^{Q_{\max}} I_c'(Q')dQ'}, \qquad (3)$$

where $I'_c(Q')$ is the scattering intensity recorded by UED. The PDF $g(r)$ is then obtained by normalizing $G(r)$:[10]

$$g(r) = \frac{G(r)}{4\pi r \rho_0} + 1, \quad (4)$$

where $\rho_0$ is the averaged density of reduced diffraction elements.